\documentclass[superscriptaddress,showpacs,aps,pra,reprint,amsmath,amssymb]{revtex4-1}                                          \usepackage{times}
\usepackage{hyperref}
\usepackage{xspace}
\usepackage{graphicx}
\usepackage{hypernat,xspace}
\usepackage[figure,table]{hypcap}
\hypersetup{colorlinks=true,
linkcolor=blue,
filecolor=blue,
citecolor=blue,
pdfkeywords={Bose-Einstein condensate, vortex, synthetic magnetic field}}

\newcommand{\im}{\ensuremath{i}}
\newcommand{\eu}{\ensuremath{e}}
\newcommand{\bvec}[1]{\ensuremath{\mathbf{#1}}}

\begin{document}
\title[Short title for running header]{Trapped Bose-Einstein condensates in synthetic magnetic field}

\author{Qiang Zhao}
\affiliation{Department of Physics, University of Science and Technology Beijing, Beijing 100083, China}
\affiliation{School of Science, North China University of Science and Technology, Tangshan 063009, China}
\author{Qiang Gu}
\email{qgu@ustb.edu.cn}
\affiliation{Department of Physics, University of Science and Technology Beijing, Beijing 100083, China}

\pacs{03.75.Lm, 03.75.Hh, 05.30.Jp}


\begin{abstract}
  Rotating properties of Bose-Einstein condensates in synthetic magnetic field are studied by numerically solving the Gross-Pitaevskii equation and compared with condensates confined in the rotating trap. It seems that it is more difficult to add large angular momentum to condensates spined up by the synthetic magnetic field than by the rotating trap. However, strengthening the repulsive interaction between atoms is an effective and realizable route to overcome this problem and can at least generate vortex-lattice-like structures. In addition, the validity of the Feynman rule for condensates in synthetic magnetic field is verified.
\end{abstract}
\maketitle

\section{Introduction}

Rotating quantum gas has attracted enormous research interest since it exhibits a number of counter-intuitive phenomena in contrast to the classical gas. One of the striking features is that vortices inside are quantized. Such features have been extensively explored in the context of superfluid $^{4}$He~\cite{Helium4} and $^{3}$He~\cite{Helium3a,Helium3b}, and superconductors~\cite{SC1994}. The recent achievement of atomic Bose-Einstein condensates (BECs) opens up a new avenue towards understanding the physics of rotating quantum gases~\cite{Fetter2009}. The atomic BEC has advantages in comparison to superfluid
$^{4}$He and $^{3}$He, since it is flexibly manipulated and can be easily rotated by means of several techniques.

The first reported vortex in BECs was created by the phase engineering technique with laser beams, in which a binary mixture of condensates was manipulated and quantized rotation of one component was realized ~\cite{Cornell1999}. Soon after, the rotating frame method succeeded in experiments, where the condensate was spined up by the rotating deformation of the confining trap~\cite{Dalibard2000a}. This approach is similar to the rotating-bucket method in experimental studies of superfluid $^{4}$He~\cite{Helium4}. One quantized vortex appeared as the rotating frequency rose to a certain critical rotating frequency of the deformed trap~\cite{Dalibard2000a,Dalibard2000b}. With the increasing of rotating frequency, the number of vortices increases larger and larger. Vortex lattices~\cite{Ketterle2001a,Ketterle2001b,Wu-Ming Liu2013} are also obtained for higher rotating frequency. Further increasing rotating frequency will drive the condensates transforms to fast rotation regime~\cite{Dalibard2004,Cornell2004}.

The angular momentum of a rotating classical fluid is proportional to its rotating frequency. For a quantum gas, the angular momentum is characterized by the total number of vortices it carries. The relation between the number of vortices and the rotating frequency is described by the well-known Feynman rule~\cite{Feynman}, which can be expressed as $ 2\pi\hbar N_V/m =2\Omega A$, where $\Omega$ is rotating frequency and $N_V$ is the vortex number within area $A$. The Feynman rule was deduced originally for the superfluid helium in the rotating-bucket. In view of the analogy between the rotating-bucket and the rotating trap, the Feynman rule is naturally applicable to the atomic BECs in the rotating trap. The validity of the Feynman rule has been intensively studied both theoretically~\cite{Masahito Ueda2002,Masahito Ueda2003,clark} and experimentally~\cite{Ketterle2001b,Haljan01} in recent years.

The rotating frame approach is subject to some limitations, e.g., it is difficult to add optical lattices and the rotation is limited by heating, metastability, etc. The synthetic magnetic field approach stands out and is expected to break above limitations~\cite{Lin2009a,Lin2009b}. This novel approach creates vector potential for atoms by dressing them in a space-dependent manner with optical field and thus makes neutral atoms behave like charged particles in the magnetic field~\cite{Lin2009a}. Vortices were observed experimentally in the condensate for synthetic magnetic field greater than the critical value~\cite{Lin2009b}. Hydrodynamical behaviors of the condensate in synthetic magnetic field and the dynamical instability of vortex nucleation were studied in Ref.~\cite{Martin}. Enormous effort was devoted to studying cold atoms subject to synthetic magnetic field in the presence of optical lattices. Experimentally, the synthetic magnetic field has been engineered in periodic lattices~\cite{Aidelsburger}. Theoretically, Berezinskii-Kosterlitz-Thouless transition in the two-dimensional lattice has been investigated~\cite{Nakano2012,Xu2012} and the Hofstadter butterfly physics in the strong field has been discussed~\cite{Lewenstein,Sols}.

Thermodynamic properties of an ideal Bose gas in the synthetic magnetic field have also been studied~\cite{fan,yushan}. It is natural to expect that rotating the gas leads to decrease of the Bose-Einstein condensation temperature. It shows that Bose-Einstein condensation can be more easily suppressed in rotating frame than in synthetic magnetic field, which implies that the rotating frame can spin up the atomic gas more efficiently than the synthetic magnetic field. If it is true, it is more difficult for the synthetic magnetic field to add large angular momentum to the condensates than the rotating frame. Therefore, it is worthwhile to check this issue carefully and to make an elaborate comparison of the two approaches.

In this paper, we focus on rotating properties of atomic BECs in the synthetic magnetic field. The vortex formation is investigated by numerically solving the Gross-Pitaevskii (GP) equation, with the emphasis on the difference between the results of the synthetic magnetic field and the rotating frame approach. This paper is organized as follows. Section~\ref{model} introduces the Gross-Pitaevskii equations of the BECs in both the magnetic field and the rotating frame. Section~\ref{DISCUSSIONS} presents the numerical results and discusses the difference between the two approaches. A brief conclusion is given in Section~\ref{SUMMARY}.

\section{The model}\label{model}

For a neutral atom of mass $m$ in synthetic magnetic field, the effective Hamiltonian is given by
\begin{align}\label{e2}
H = \frac{1}{2m}(\boldsymbol{P} - \boldsymbol{A})^2
    + \frac{1}{2}m\omega_0^2 (x^2+y^2)+\frac{1}{2}m \omega^2_z z^2 ,
\end{align}
where $\boldsymbol{P}$ is the canonical momentum operator and $\boldsymbol{A}$ is the synthetic gauge potential. Here the atomic Bose-Einstein condensate is confined in the anisotropic harmonic trap, and $\omega_0$ and $\omega_z$ are the trap frequencies in the $x$-$y$ plane and the $z$ axis. We assume that the synthetic field $\mathbf{B}=2m\Omega\vec{e}_z$ is parallel to the $z$ axis and choose the symmetric gauge $\mathbf{A}=(\mathbf{B}\times\mathbf{r})/2=m(\mathbf{\Omega}\times\mathbf{r})$, where $\Omega$ is the Larmor frequency. Therefore, the GP equation for a Bose-Einstein condensate can be written as
\begin{align}\label{e3b}
\im\hbar\frac{\partial \psi(\bvec r, t)}{\partial t}= \left[ \frac{1}{2m}(\boldsymbol{P} - \boldsymbol{A})^2 + V^\prime(\bvec r) + g|\psi|^2   \right] \psi(\bvec r,t),
\end{align}
where $V^\prime(\bvec r)=\frac{1}{2}m\omega_0^2(x^2+y^2)+\frac{1}{2}m \omega^2_z z^2$, $\psi(\bvec r,t)$ is the wave function of the condensates, and $g = 4\pi\hbar^2 a_s/m$ is the contact interaction strength between atoms with s-wave scattering length $a_s$. In the present work, we consider pancake-shaped BECs, that is, the harmonic trapping frequencies satisfy $\omega _{z}\gg \omega_0$.

For convenience, Eq.~\eqref{e3b} can be expressed with dimensionless quantities, where the spatial coordinates $x$, $y$ and $z$ are normalized by the characteristic harmonic oscillator length $a_0 = \sqrt{\hbar/(m\omega)}$ with $\omega =\rm min~ \{\omega_0,\omega_z\}=\omega_0$ and the time $t$ is in units of $\omega_0^{-1}$. Then the Eq.~\eqref{e3b} is reduced to a dimensionless form
\begin{align}
i\frac{\partial\tilde{\psi}(\tilde{\bvec{r}},\tilde{t})}{\partial\tilde{t}}
  &= \left[ \frac{1}{2}(\boldsymbol{\tilde{P}} - \boldsymbol{\tilde{A}})^2 + V(\tilde{\bvec r}) + \beta |\tilde{\psi}|^2  \right]\tilde{\psi}(\tilde{\bvec{r}},\tilde{t})\label{e4_b},
\end{align}
where $V(\tilde{\bvec r})= \frac{1}{2}\gamma_0^2( \tilde{x}^2 +\tilde{y}^2) +\frac{1}{2}
\gamma_z^2 \tilde{z}^2$, $\gamma_0 = \omega_0/\omega=1$, $\gamma_z = \omega_z/\omega_0$, $\beta = 4\pi N a_s/a_0$. Here, dimensionless variables are denoted with a tilde.
Since $\omega _{z}\gg \omega_0$, the wave function can be supposed to be in a
variables separation form
$$\tilde{\psi}(\tilde{\bvec{r}},\tilde{t}) = \tilde{\psi}_\text{2D}(\tilde{x}, \tilde{y}, \tilde{t}) \tilde{\psi}_\text{1D}(\tilde{z}) \eu^{-\im\gamma_z \tilde{t}/2},$$
where
$$\tilde{\psi}_\text{1D}(\tilde{z}) = (\gamma_z/\pi)^{1/4}\eu^{-\gamma_z \tilde{z}^2/2}$$    
is the ground state of the harmonic oscillator along the $z$ direction. After integrating out the
coordinate $z$, we obtain the two-dimensional (2D) GP equation
\begin{align}\label{e-syn}
i\frac{\partial\tilde{\psi}_\text{2D}(\tilde{x},\tilde{y},\tilde{t})}{\partial\tilde{t}}=& \left[ \frac{1}{2}(\boldsymbol{\tilde{P}} - \boldsymbol{\tilde{A}})^2+ \frac{1}{2}( \tilde{x}^2 +\tilde{y}^2) \right.\nonumber\\
 &\left.+g_\text{2D}|\tilde{\psi}_\text{2D}|^2  \right] \tilde{\psi}_\text{2D}(\tilde{x}, \tilde{y}, \tilde{t}) ,
\end{align}
where $g_\text{2D}=\beta\sqrt{\gamma_z/2\pi}$ represents the effective 2D interaction strength. We note that $\gamma_z$ is just the trap aspect ratio of the anisotropic trap.

For comparison, we also consider properties of BECs in the rotating frame. In this case, the single-particle Hamiltonian is given by
\begin{align}\label{e1}
H = \frac{1}{2m}(\boldsymbol{P} - \boldsymbol{A})^2
    +\frac{1}{2m}(\omega^2_0-\Omega^2)(x^2+y^2)+
    \tfrac{m}{2}\omega_z^2 z^2.
\end{align}
Here $\mathbf{A}=m(\mathbf{\Omega}\times\mathbf{r})$ and $\Omega$ denotes the rotating frequency of the trap which is comparable to the Larmor frequency in Eq.~\eqref{e2}. Therefore, both of them are referred as the rotating frequency in this paper. With the same procedure, we have the corresponding 2D GP equation in the dimensionless form
\begin{align} \label{e-rot}
i\frac{\partial\tilde{\psi}_\text{2D} (\tilde{x},\tilde{y},\tilde{t})}{\partial\tilde{t}}=& \left[ \frac{1}{2}(\boldsymbol{\tilde{P}} - \boldsymbol{\tilde{A}})^2
+ \frac{1}{2} (1-\frac{\Omega^2}{\omega^2_0})(\tilde{x}^2 +\tilde{y}^2)\right.\nonumber\\
 &\left.+ g_\text{2D}|\tilde{\psi}_\text{2D}|^2 \right] \tilde{\psi}_\text{2D}(\tilde{x}, \tilde{y}, \tilde{t}).
\end{align}

In our calculation, we use the Fourier spectral method to solve the nonlinear differential Eqs.~\eqref{e-syn} and \eqref{e-rot} via the imaginary time propagation approach~\cite{bao2006}. Although these two
equations have no fundamental difference, the comparison indicates
apparent different physics behaviour. The calculation starts with some initial states, then propagates until numerical convergence is achieved.

\begin{figure}
\begin{center}
\setlength{\unitlength}{1cm}
\includegraphics[width=8cm,angle=0]{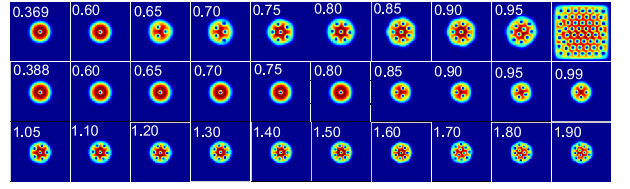}
\end{center}
\caption{Contour plots of the density distributions $|\psi_\text{2D}|^2$ showing a steady vortex state in a rotating BECs with the dimensionless contact interaction $g_\text{2D}=100$. Upper panels correspond to rotating frame. The middle (lower) panels correspond to synthetic magnetic field. The number in the panels shows the rotating frequency in units of $\omega_0$. The last figure in the upper panels corresponds to rotating frequency $0.99 \omega_0$. The field of view in the panels is $8a_0\times 8a_0$.}
\label{fig1}
\end{figure}

\section{Results and discussions} \label{DISCUSSIONS}

The Gross-Pitaevskii equation provides a remarkably reliable description of the physics of the atomic condensate. Properties of atomic BECs in the rotating frame have been intensively investigated based on the GP equation~\cite{Masahito Ueda2002,Masahito Ueda2003,clark}. The formation of vortices with relation to the rotating frequency, the contact interaction between atoms and the trap aspect ratio of traps have been discussed in detail. Hereinafter, corresponding results for the BECs in synthetic magnetic field are examined. The emphasis is laid on the difference between the two cases.

\subsection{Vortex formation with relation to rotating frequency}

The formation of quantum vortices and the dependence of the vortex number on the rotating frequency are among the most important issues for rotating quantum gases. Different from the classical case, no angular momentum can be added to the quantum gas until the rotating frequency exceeds a critical value, when one quantize vortex begins to be created. Then more and more vortices appear as further increases the rotating frequency.

Figure~\ref{fig1} shows the two-dimensional atom density at different rotating frequencies for BECs with the dimensionless contact interaction $g_\text{2D}=100$ in the rotating frame and synthetic magnetic field. The critical rotating frequency of creating the first vortex inside the BECs in the rotating frame and synthetic magnetic field is $0.369\omega_0$ and $0.388\omega_0$, respectively. The two values are very similar and the latter is slightly larger than the former.

However, the two rotating approaches take on remarkable difference in high rotating frequency cases. For BECs in the rotating frame, the vortex number grows quickly with $\Omega$, especially when $\Omega$ approaches the trapping frequency $\omega_0$. For example, there are about $10$ vortices at $\Omega/\omega_0=0.95$ and the vortex number amounts to 56 at $\Omega=0.99\omega_0$ which is just the upper limit of the rotating frequency. Meantime, the BEC expands significantly. It can be seen that the condensate occupies almost the whole square region under consideration in our calculation when $\Omega$ rises to $0.99\omega_0$. The vortex structure looks like the Abrikosov vortex lattice. We note that the obtained results may not be sufficiently accurate when $\Omega$ is close to $\omega_0$, because the boundary condition we choose in the calculation may not be well met.

For the BECs rotated by the synthetic magnetic field, it is obvious that the vortex number grows much more slowly. After one vortex is created, the vortex number keeps unchanged even if $\Omega$ reaches $0.8\omega_0$. It seems rather difficult to add more vortices. Only $4$ vortices emerge at $\Omega=0.99\omega_0$. The Larmor frequency can be larger than $\omega_0$. Actually, there is no intrinsic upper limit value for $\Omega$ and it only depends on the strength of the synthetic magnetic field. Figure~\ref{fig1} shows the results for $\Omega>\omega_0$. The vortex number rises, but still in a slow way. There are about 12 vortices at $\Omega=1.9\omega_0$. The number is still small. In experiments, about 10 vortices were observed in the BECs under the synthetic field~\cite{Lin2009a,Lin2009b}.

Another difference from the rotating frame case is that the size of the condensates maintains almost unchanged as $\Omega$ increases. This point is clearly shown in Fig.~\ref{fig1}.

\begin{figure}[htb]
\centering
\includegraphics[width=0.40\textwidth]{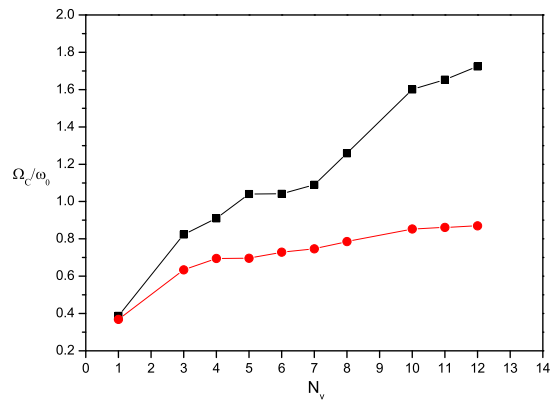}
\caption{ Plot of the equilibrium vortex number $N_V$ and rotating frequency of creating certain number vortices $\Omega_c/\omega_0$ for conventional BECs with the dimensionless contact interaction $g_\text{2D}=100$. The square points and circle points are numerical results. The black and red lines represent synthetic magnetic field and rotating frame, respectively.}
\label{fig2}
\end{figure}

\begin{figure}[htb]
\centering
\includegraphics[width=0.40\textwidth]{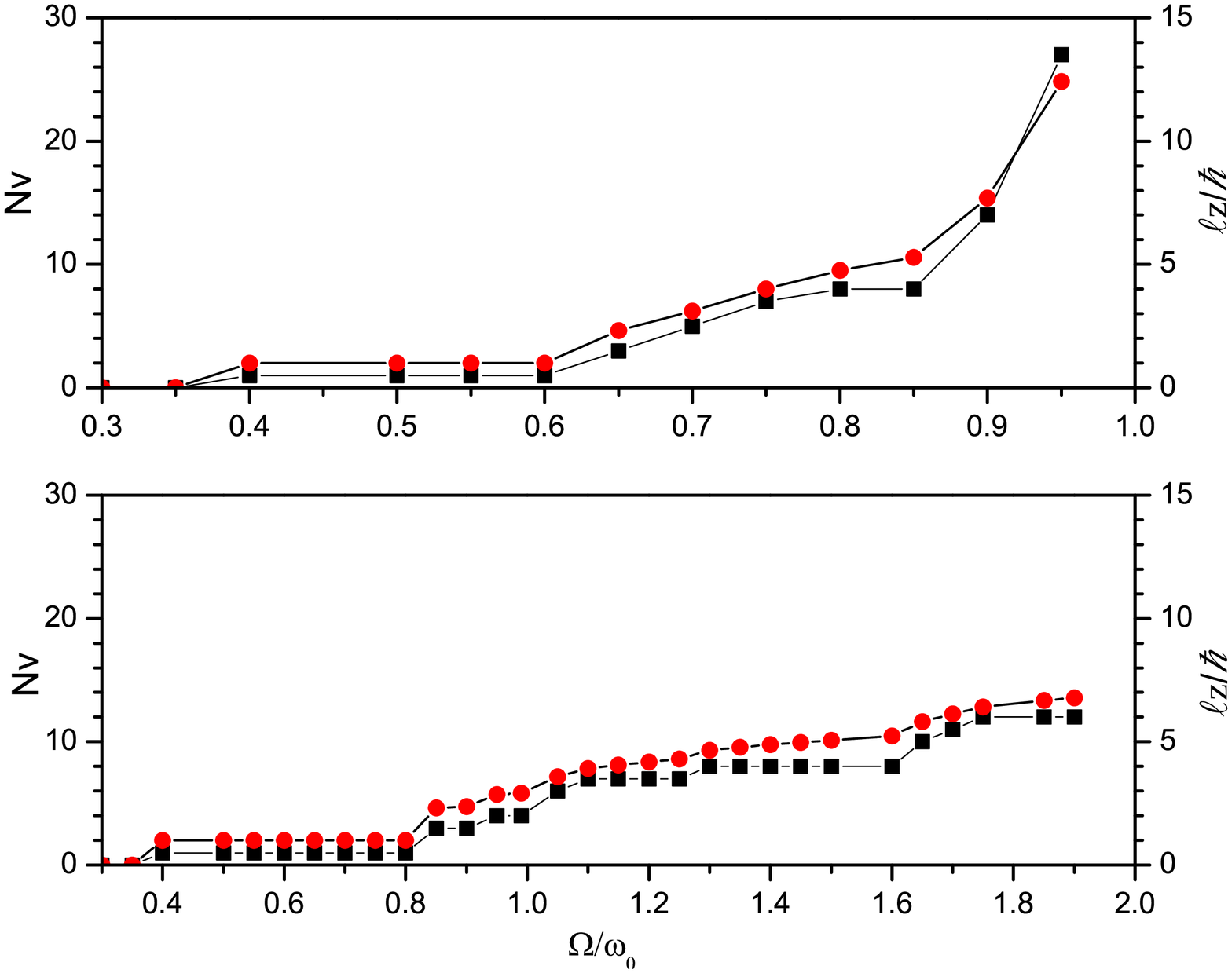}
\caption{The number of vortices $N_V$ (black square points) and angular momentum per atom $\ell_\mathrm{z}/\hbar$ (red circle points) versus rotating frequency $\Omega/\omega_0$ for conventional BECs with the dimensionless contact interaction $g_\text{2D}=100$. The upper and bottom panels correspond to rotating frame and synthetic magnetic field, respectively.}
\label{fig3}
\end{figure}

Above results indicate that it is hard to create vortex in BECs rotated by the synthetic magnetic field, satisfying the indication of thermodynamic calculations~\cite{fan,yushan}. To illustrate this point further, we calculate the critical rotating frequencies, $\Omega_c/\omega_0$, of creating certain number of vortices, $N_V$, for the two rotating cases. The results are shown in Fig.~\ref{fig2}. As mentioned above, in two cases, the critical rotating frequencies for creating one vortex are very similar. As the vortex number increases, the two corresponding critical rotating frequencies become more and more different. For example, in order to create twelve vortices, the critical rotating frequency in the synthetic field case is about twice as large as that in rotating frame case.

It should be noted that the vortices are admitted into the condensate via the dynamically instability~\cite{Subhasis,Martin} and if the dynamical process of seeding vortices inside the condensate is taken into account the critical rotating frequency should be larger than the value obtained within the present Gross-Pitaevskii equation approach. Nevertheless, as analyzed above our treatment can provide a  qualitatively reasonable description of the vortex formation process.

The Feynman rule for atomic BECs in the rotating trap has been studied~\cite{Masahito Ueda2002,Masahito Ueda2003,clark,Ketterle2001b,Haljan01}. To proceed, we assess the validity of the Feynman rule for the BECs in synthetic magnetic field. Figure~\ref{fig3} shows the dependence of the vortex number $N_V$ and the angular momentum per atom $\ell_\mathrm{z}/\hbar$ on the rotating frequency $\Omega/\omega_0$. The number $N_V$ is counted from the density distribution and the angular momentum per atom is numerically calculated according to the equation $\ell_\mathrm{z}=\iint\psi^{*}L_z\psi dxdy/\iint|\psi|^2 dxdy$, where $L_z=-i\hbar \left( x\partial _{y}-y\partial _{x}\right)$ is the $z$-component of the angular momentum operator. Numerical results show that the angular momentum per atom $\ell_\mathrm{z}/\hbar$ is about a half of the number of vortices, the Feynman rule is basically met for the two frames. The small disagreement between $N_V$ and $\ell_\mathrm{z}/\hbar$ may be attributed to the inhomogeneous density~\cite{Masahito Ueda2002}.

\subsection{Vortex formation with relation to contact interaction}

\begin{figure}[htb]
\centering
\includegraphics[width=0.40\textwidth]{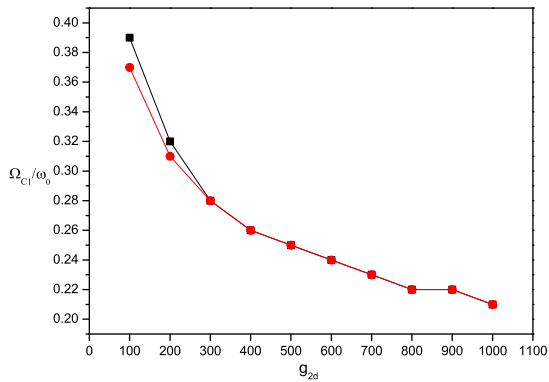}
\caption{The lower critical rotating frequency $\Omega_{c1}/\omega_0$ for the single-vortex state as a function of the dimensionless contact interaction $g_\text{2D}$. The square points and circle points are numerical results. The black and red lines represent synthetic magnetic field and rotating frame, respectively.}
\label{fig4}
\end{figure}

According to the investigation on BECs in the rotating frame, the interaction between atoms can affect the formation of vortices considerably~\cite{anjin}. Vortices can be created more easily in the BECs with stronger repulsive interactions. It is of interest to study the role of the contact interaction for BECs in the synthetic field. Figure~\ref{fig4} plots the critical rotating frequency $\Omega_{c1}/\omega_0$ for the single-vortex state as a function of the dimensionless contact interaction $g_\text{2D}$. The lower critical rotating frequency decreases monotonically with the contact interaction for both rotating approaches, which indicates that the interaction results in similar effect on the formation of vortices in BECs rotated by the synthetic magnetic field.

\begin{figure}[htb]
\centering
\includegraphics[width=0.40\textwidth]{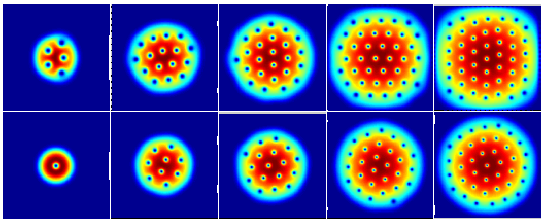}
\caption{Upper and bottom panels: density profiles
for rotating frame and synthetic magnetic field at $\Omega=0.7\omega_0$. From left to right, the dimensionless contact interaction is 100, 500, 1000, 2000 and 3000, respectively. The field of view in the panels is $8a_0\times 8a_0$.}
\label{fig5}
\end{figure}

Figure~\ref{fig5} illustrates vortex structures at different values of $g_\text{2D}$. The rotating frequency is relatively large, $\Omega=0.7\omega_0$. The upper and bottom panels show the atom density for the rotating frame and synthetic field cases, respectively. By simply counting, we find the ratio of vortex numbers between the two cases are 5/1, 16/9, 24/17, 39/28, and 46/39 at $g_\text{2D}=100$, $500$, $1000$, $2000$ and $3000$, respectively. For both cases, the vortex number increases with increasing the interaction, and the ratio decreases and tends to $1$. These results suggest that although it is still relatively difficult to generate vortices in the latter case in comparison to the former one, the difference between two cases becomes small at sufficiently large contact interactions. For example, vortex numbers are comparable at $g_\text{2D}=3000$ and lattice-like structures are formed in two cases. The vortex lattice hasn't been observed in experiments by the latter approach~\cite{Lin2009b}. In atomic gases, the s-wave scattering length can be tuned easily via the Feshbach resonance~\cite{Feshbach1,Feshbach2}. So increasing of contact interactions is an effective and realizable route to add more angular momentum and generate more vortices in the condensates rotated by the synthetic magnetic field.

As shown in Fig.~\ref{fig5}, the condensate size is enlarged apparently with strengthening the interaction. It appears that it is easier to add vortices in a condensate of larger size when the particle number keeps constant. This insinuate the possible reason why it is easy to produce more vortices in the BECs spined up by the rotating trap than by the synthetic magnetic field. This might be due to the confinement of trap potential. According to Eqs.~\eqref{e-syn} and \eqref{e-rot}, in the rotating frame, the effective trapping potential is weakened with the rotating frequency $\Omega$, thus the condensate expands correspondingly. Nevertheless, in the synthetic magnetic field case, the trapping potential is independent on the Larmor frequency.

At last, we briefly discuss the effect of the trap aspect ratio, $\gamma_z =\omega_z/\omega_0$, on the formation of vortices. This issue has been discussed for 3-dimensional condensates in the rotating frame and it is indicated that increasing $\gamma_z$ helps nucleation of the vortex~\cite{Lewenstein2000}. In the present study the 3D system is converted into an effective 2D model and the trap aspect ratio $\gamma_z$ is included in the effective 2D contact interaction $g_\text{2D}=\beta\sqrt{\gamma_z/2\pi}$, the effect of increasing $\gamma_z$ is similar to that of increasing $g_\text{2D}$, as shown in Fig.~\ref{fig5}.

\section{Summary} \label{SUMMARY}

We have investigated rotating properties of the Bose-Einstein condensates in synthetic magnetic field. The formation of vortices are calculated by numerically solving the Gross-Pitaevskii equation, considering effects of the Larmor (rotating) frequency, the interaction between atoms and the trap aspect ratio. The obtained results are compared with those of the BECs in the rotating frame. We find that the vortex number in the condensate spined up by the synthetic magnetic field is much smaller than in that by the rotating frame given the same rotating frequency, which implies that the synthetic magnetic field has less efficiency in rotating the condensate. Strengthening the repulsive interaction between atoms or increasing the trap aspect ratio are helpful for creating more vortices. In particular, when the interaction is sufficiently strong, comparable large number of vortices are produced by both rotating approaches. Abrikosov lattice-like structures can be formed in two cases. Mover, the validity of the Feynman rule is checked for the BECs in synthetic magnetic field case.

\begin{acknowledgements}

The authors are grateful to Weizhu Bao for valuable assistance in the numerical and programming techniques. This work is supported by the National Key Basic Research Program of China (Grant No. 2013CB922002), the National Natural Science Foundation of China (Grant No. 11074021) and the Fundamental Research Funds for the Central Universities of China.

\end{acknowledgements}



\begin{thebibliography}{99}

\bibitem{Helium4} R. J. Donnelly, Quantum vortices in Helium II, Cambridge: Cambridge University Press, 1991

\bibitem{Helium3a} D. Vollhardt and P. W\"{o}lfle, The superfluid phases of Helium 3, London: Taylor \& Francis, 1990

\bibitem{Helium3b}G. E. Volovik, The universe in a Helium droplet, Oxford: Clarendon, 2003

\bibitem{SC1994} G. Blatter, M. V. Feigel'man, V. B. Geshkenbein, A. I. Larkin, and V. M. Vinokur, Vortices in high-temperature superconductors, \emph{Rev. Mod. Phys.} 66(4), 1125 (1994)

\bibitem{Fetter2009} A. L. Fetter, Rotating trapped Bose-Einstein condensates, \emph{Rev. Mod. Phys.} 81(2), 647 (2009)

\bibitem{Cornell1999} M. R. Matthews, B. P. Anderson, P. C. Haljan, D. S. Hall, C. E. Wieman, and E. A. Cornell, Vortices in a Bose-Einstein condensate, \emph{Phys. Rev. Lett.} 83(13), 2498 (1999)

\bibitem{Dalibard2000a} K. W. Madison, F. Chevy, W. Wohlleben, and J. Dalibard, Vortex formation in a stirred Bose-Einstein condensate, \emph{Phys. Rev. Lett.} 84(5), 806 (2000)

\bibitem{Dalibard2000b} F. Chevy, K. W. Madison, and J. Dalibard, Measurement of the angular momentum of a rotating Bose-Einstein condensate, \emph{Phys. Rev. Lett.} 85(11), 2223 (2000)

\bibitem{Ketterle2001a} C. Raman, J. R. Abo-Shaer, J. M. Vogels, K. Xu, and W. Ketterle, Vortex nucleation in a stirred Bose-Einstein condensate, \emph{Phys. Rev. Lett.} 87(21), 210402 (2001)

\bibitem{Ketterle2001b} J. R. Abo-Shaeer, C. Raman, J. M. Vogels, and
W. Ketterle, Observation of vortex lattices
in Bose-Einstein condensates, \emph{Science} 292(5516), 476 (2001)

\bibitem{Wu-Ming Liu2013} S.-W. Song, L. Wen, C.-F. Liu, S.-C. Gou, and W.-M. Liu, Ground states, solitons and spin textures in spin-1 Bose-Einstein condensates, \emph{Frontiers of physics} 8(3), 302-318 (2013)

\bibitem{Dalibard2004} V. Bretin, S. Stock, Y. Seurin, and J. Dalibard, Fast rotation of a Bose-Einstein condensate, \emph{Phys.
Rev. Lett.} 92(5), 050403 (2004)

\bibitem{Cornell2004} V. Schweikhard, I. Coddington, P. Engels, V. P. Mogendorff, and E. A. Cornell, Rapidly rotating Bose-Einstein condensates in and near the lowest Landau level, \emph{Phys. Rev. Lett.} 92(4), 040404 (2004)

\bibitem{Feynman} R. P. Feynman, Application of quantum mechanics to liquid Helium, Amsterdam: North-Holland, 1955

\bibitem{Masahito Ueda2002} M. Tsubota, K. Kasamatsu, and M. Ueda, Vortex lattice formation in a rotating Bose-Einstein condensate, \emph{Phys. Rev. A} 65(2), 023603 (2002)

\bibitem{Masahito Ueda2003} K. Kasamatsu, M. Tsubota, and M. Ueda, Nonlinear dynamics of vortex lattice formation in a rotating Bose-Einstein condensate, \emph{Phys. Rev. A} 67(3), 033610 (2003)

\bibitem{clark} D. L. Feder and C. W. Clark, Superfluid-to-Solid crossover in a rotating Bose-Einstein condensate, \emph{Phys. Rev. Lett.} 87(19), 190401 (2001)

\bibitem{Haljan01} P. C. Haljan, I. Coddington, P. Engels, and E. A. Cornell, Driving Bose-Einstein-Condensate vorticity with a rotating normal cloud, \emph{Phys. Rev. Lett.} 87(21), 210403 (2001)

\bibitem{Lin2009a} Y.-J. Lin, R. L. Compton, A. R. Perry, W. D. Phillips, J. V. Porto, and I. B. Spielman, Bose-Einstein condensate in a uniform light-induced vector potential, \emph{Phys. Rev. Lett.} 102(13), 130401 (2009)

\bibitem{Lin2009b} Y.-J. Lin, R. L. Compton, K. Jim\'{e}nez-Garc\'{i}a, J. V. Porto and I. B. Spielman, Synthetic magnetic fields for ultracold neutral atoms, \emph{Nature} 462(7273), 628 (2009)

\bibitem{Martin} L. B. Taylor, R. M. W. van Bijnen, D. H. J. O'Dell, N. G. Parker, S. J. J. M. F. Kokkelmans, and A. M. Martin, Synthetic magnetohydrodynamics in Bose-Einstein condensates and routes to vortex nucleation, \emph{Phys. Rev. A} 84(2), 021604(R) (2011)

\bibitem{Aidelsburger} M. Aidelsburger, M. Atala, S. Nascimbene, S. Trotzky, Y.-A. Chen, and I. Bloch , Experimental realization of strong effective magnetic fields in an optical lattice, \emph{Phys . Rev. Lett.} 107(25), 255301 (2011)

\bibitem{Nakano2012} Y. Nakano, K. Kasamatsu, and T. Matsui, Finite-temperature phase structures of hard-core bosons in an optical lattice with an effective magnetic field, \emph{Phys. Rev. A} 85(2), 023622 (2012)

\bibitem{Xu2012} J. Xu and Q. Gu, Berezinskii-Kosterlitz-Thouless transition of two-dimensional Bose gases in a synthetic
magnetic field, \emph{Phys. Rev. A} 85(4), 043608 (2012)

\bibitem{Lewenstein} A. Celi, P. Massignan, J. Ruseckas, N. Goldman, I. B. Spielman, G. Juzeli\=unas, and M. Lewenstein, Synthetic gauge fields in synthetic dimensions, \emph{Phys. Rev. Lett.} 112(4), 043001 (2014)

\bibitem{Sols} C. E. Creffield and F. Sols, Generation of uniform synthetic magnetic fields by split driving of an optical lattice, \emph{Phys. Rev. A} 90(2), 023636 (2014)

\bibitem{fan} J.-H. Fan, Q. Gu, W. Guo, Thermodynamics of charged ideal Bose gases in a trap under a magnetic Field, \emph{Chin. Phys. Lett.} 28(6), 060306 (2011)

\bibitem{yushan} Y. Li and Q. Gu, Thermodynamic properties of rotating trapped ideal Bose gases, \emph{Phys. Lett. A} 378(18-19), 1233 (2014)

\bibitem{bao2006} W. Bao, I-L. Chern, and F. Y. Lim, Efficient and spectrally accurate numerical methods
for computing ground and first excited states in
Bose-Einstein condensates, \emph{J. Comp. Phys.} 219(2), 836 (2006)

\bibitem{Subhasis} S. Sinha and Y. Castin, Dynamic Instability of a Rotating Bose-Einstein Condensate, \emph{Phys. Rev. Lett.} 87(19), 190402 (2001)

\bibitem{anjin} Y. Zhao, J. An, and C.-D. Gong, Vortex competition in a rotating two-component dipolar Bose-Einstein condensate, \emph{Phys. Rev. A} 87(1), 013605 (2013)

\bibitem{Feshbach1} G. Thalhammer, G. Barontini, L. De Sarlo, J. Catani, F. Minardi, and M. Inguscio, Double species Bose-Einstein condensate with tunable interspecies interactions, \emph{Phys. Rev. Lett.} 100(21), 210402 (2008)

\bibitem{Feshbach2} C. Chin, R. Grimm, P. Julienne, and E. Tiesinga, Feshbach resonances in ultracold gases, \emph{Rev. Mod. Phys.} 82(2), 1225 (2010)

\bibitem{Lewenstein2000} L. Santos, G. V. Shlyapnikov, P. Zoller and M. Lewenstein, Bose-Einstein condensation in trapped dipolar gases, \emph{Phys. Rev. Lett.} 85(9), 1791 (2000)

\end{thebibliography}
\end{document}